\documentclass[12pt,showpacs,nofootinbib]{revtex4-1}
\usepackage{multirow}
\usepackage{amssymb}
\usepackage{tikz}
\usepackage{amssymb}
\usepackage{amsmath,graphicx}
\usepackage[compat=1.1.0]{tikz-feynman}
\usepackage{array}
\usepackage{dcolumn}
\usepackage{braket}
\usepackage{bm}
\usepackage{enumerate}
\usepackage{slashed}
\usepackage{epstopdf}
\usepackage{amsmath}
\usepackage{dcolumn,caption,booktabs}
\usepackage{slashed}
\usepackage{amsfonts}
\usepackage{adjustbox}
\usepackage{tcolorbox}
\usepackage{float}
\usepackage{collectbox}
\usepackage{listings}
\usepackage[unicode]{hyperref}
\usepackage[mathscr]{eucal}
\usepackage{graphics}
\usepackage{subcaption}
\newcommand{\be}{\begin{equation}}
	\newcommand{\ee}{\end{equation}}
\newcommand{\bea}{\begin{eqnarray}}
	\newcommand{\eea}{\end{eqnarray}}
\newcommand{\ben}{\begin{enumerate}}
	\newcommand{\een}{\end{enumerate}}
\newcommand{\bde}{\begin{widetext}}
	\newcommand{\ede}{\end{widetext}}

\newcommand{\crn}{\nonumber \\}

\newcommand{\al}{\alpha}

\newcommand{\fr}{\frac}

\newcommand{\bc}{\begin{center}}
	\newcommand{\ec}{\end{center}}

\newcommand{\ep}{\epsilon}

\newcommand{\Om}{\Omega}

\newcommand{\betah}{\frac{\beta}{H*}}
\begin{document}
	\newcommand{\AdrHEPC}{$^a$Department of Theoretical Physics, University of Science, Ho Chi Minh City 70000, Vietnam\\ $^b$Vietnam National University, Ho Chi Minh City 70000, Vietnam}
	
	\title{Twin electroweak bubble nucleation and gravitational wave under the $S_3$ symmetry of two-Higgs-doublet model}
	\author{Vo Quoc Phong$^{a,b}$}
	\email{vqphong@hcmus.edu.vn}
	\affiliation{\AdrHEPC}
	\author{Nguyen Xuan Vinh$^{a,b}$}
	\email{vinhnguyen.mxt@gmail.com }
	\affiliation{\AdrHEPC}
	\author{Phan Hong Khiem$^{c,d}$}
	\email{phanhongkhiem@duytan.edu.vn}
	\affiliation{$^c$Institute of Fundamental and Applied Sciences, Duy Tan University, Ho Chi Minh City 70000, Vietnam\\
		$^d$ Faculty of Natural Sciences, Duy Tan University, Da Nang City 50000, Vietnam}	
\begin{abstract}
Sphaleron electroweak phase transition (EWPT) is calculated in two phase transition stages, thereby showing that the twin (or double) bubble nucleation structure of the phase transition and gravitational wave is in the investigation area of future detectors. With $v^2=v^2_1+v^2_2$ ($v_1$ and $v_2$ are two vacuum average values (VEV)) and $a=v^2/v_2^2$ which affects the expansion of bubbles during two phase transitions. The more $a$ increases, the more the expansion of two bubbles is at the same time. This ratio does not greatly affect the sphaleron energy but has an impact on gravitational waves. The larger the masses of the charged Higgs particles are, the greater the gravitational wave energy density ($\Omega h^2$) is. When the frequency is in the range $0-1.2$ mHz, $\Omega h^2$ will has a maximum value in the range $10^{-12}-10^{-11}$ for all values of $a$ so this can be detected in the future.
\end{abstract}
	\maketitle	

\tableofcontents
	
\section{INTRODUCTION}\label{secInt}
	
Particle physics today not only studies each individual problem in depth but also combines it with Cosmology and Astronomy. The baryon asymmetry is a problem of intersection among those fields. The possible causes of this problem are summarized and presented quite extensively. Indeed, for Higgs 125 GeV, Standard Model (SM) has only a second order electroweak phase transition when the mass of Higgs is greater than 75 GeV \cite{mkn,SME,SMEb,SMEc,SMEd,michela}. Therefore, SM should be extended (for example, Refs.~\cite{plv,2b,2c,BSM,BSMb,majorana,majoranab,thdmb,ESMCO,elptdm,elptdma,elptdmb, elptdmc,elptdmd, phonglongvan,phonglongvanb,phonglongvan2,SMS,munusm,lr,singlet,singletb,singletc,singletd,mssm1,mssm1b,mssm1c,twostep,twostepb,twostepc,1101.4665,1101.4665b,1101.4665c,jjgb,jjgc, jjgd, scaling1, scaling2, scaling3, Ahriche3, Ahriche3b, Ahriche3c, Ahriche3d, Ahriche3e, chr, cde, kusenko}). There are 3 main branches: the first is due to the existence of exotic particles (beyond the standard model (SM)); the second is because of parameters or interactions in SM changing at a electroweak transition temperature \cite{plv,2b,2c,BSM,BSMb,majorana,majoranab,thdmb,ESMCO,elptdm,elptdma,elptdmb,elptdmc,elptdmd,phonglongvan,phonglongvanb,phonglongvan2,SMS,munusm,lr,singlet,singletb,singletc,singletd,mssm1,mssm1b,mssm1c,twostep,twostepb,twostepc,1101.4665,1101.4665b,1101.4665c,jjgb,jjgc,jjgd,scaling1,scaling2,scaling3,Ahriche3b,Ahriche3c,Ahriche3d,Ahriche3e,chr,cde,kusenko}; the third be having more than 4 dimensions of spacetime \cite{pn}.

Except for the Leptongenesis scenario, the more familiar one for the baryon asymmetry is Baryogenesis. The center condition is a first order electroweak phase transition \cite{sakharov}, is needed to sufficiently maintain a thermal imbalance at a critical temperature during the mass generation of elementary particles. This condition down to calculating the high-temperature effective Higgs potential and investigating the electroweak phase transition (EWPT).

Two scenarios are commonly seen when analyzing the structure of EWPT process: the single-stage scenario as in Refs. \cite{zb,pkll,yukawa1,yukawa2, gww}, most theories under this scenario are models that modify SM; and mixed or separated multi-phase scenarios \cite{plv,phonglongvan2,ptl,pa}, via models have group structures different from SM.

Currently there are three methods to survey the effective potential of EWPT process. The first method: The effective potential $V_{eff}(v)$ which is a function of $v$ (VEV), written jointly for two VEVs and $v^2=v^2_1+v_2^2$. The second one in Refs.~\cite{dori,mayumi}: the effective potential is a function of two VEVs but EWPT is still one phase as SM. The third method is ours, in Ref.\cite{pal}: the effective potential into two separate parts $[V_{eff}=V_{eff}(v_1)+V_{eff}(v_2)]$ and examining two discrete stages. These methods all require the effective potential. However, considering the mixing components of VEVs will lead to whether the electroweak phase transition is a single-stage or a multi-stage process. 

The two-Higgs-doublet model with $S_3$ symmetry (2HDM-$S_3$) \cite{1601}, an extended version of 2HDM. 2HDM-$S_3$ has many new particles and two VEVs, with suitable structures and triggers for multi-stage electroweak phase transition calculations as well as sphaleron and GW analysis, and it has been studied by us in terms of triggers for a first order electroweak phase transition \cite{pal}. Other roles of $S_3$ symmetry in the quark sector or the Yukawa couplings, are discussed in Refs. \cite{bpal1,bpal2,bpal3,braconi}.

The first order EWPT was solved in the two-Higgs-doublet model with $S_3$ symmetry. When adding the $S_3$ symmetry to 2HDM, it makes the path of EWPT process discrete. This decoherence is expressed by two discretization stages corresponding to two VEVs. Typically, the first phase goes with $v_2$, the second phase goes with $v_1$.	
	\begin{align*}
		\centering
		&\text{2HDM-$S_3$}&\\
		&\Downarrow \text{breaking $v_2$}\\
		&\Downarrow \text{breaking $v_1$} \\
		&U(1)_Q
	\end{align*}

The $S_3$ symmetry acts as the one to simplify the parameters of Higgs potential and can make the EWPT process discrete. But it does not play the dominant role in terms of particle interaction dynamics in 2HDM. Therefore, the value domains of $\tan\beta=v_2/v_1$ drawn from different data in 2HDM are still valid in 2HDM-$S_3$. These remarks are like the following conclusions:
\begin{itemize}	
	\item According to Ref.~\cite{beta1,beta2,bpal1,bpal2,dori}, when calculating decay channels and combining LHC data, $1<\tan\beta=v_2/v_1<17$. Cases of $\tan\beta$ less than one do not exist. The masses of heavy particles are more than 190 GeV\cite{mayumi,CMS2,126}.
	
	\item On the other hand, the independent gauge of EWPT (specifically the effective potential and the phase transition strength) \cite{zb, 1101.4665, 1101.4665b,1101.4665c,Arefe}. The daisy loops are not the main cause of EWPT but reduces the strength of EWPT \cite{r23}. Therefore, when investigating around the electroweak phase transition temperature region, for simplicity we can ignore their contributions.
	
	\item In Refs. \cite{1305.6610, 1504.05949, PRL,125}, the strong first order EWPT in 2HDM with the condition, the hierarchy of mass, $m_{H^\pm }<m_{H} < m_{A}$ and $1 < \tan\beta <10$.
	
	\item The results in Refs.~\cite{2111.13079J. High Energy Phys,128} confirm that the masses of additional Higgs bosons are typically $300-400$ GeV for a first order EWPT. 
	
	\item Furthermore, in Ref.~\cite{dori,124,127,kanemura,thomas} there are many scenarios among the additional Higgs bosons were studied in the EWPT problem, specifically $m_{H^\pm}=m_A$ or $m_{H}=m_{H^\pm }$.	
\end{itemize}

The above conclusions lead to an instruction for surveying 2HDM-$S_3$ also with a parameter region, $1\sim\tan\beta<17$ and the masses of additional bosons must be larger than 200 GeV. Thus, the value domain of $\tan\beta$ is usually quite wide, which is not convenient for investigating EWPT. Therefore, following Ref.\cite{pal}, we introduce the parameter $a=v^2/v^2_2$ to replace $\tan\beta$. This will also be specified again in Section II. Our result in Ref.\cite{pal} agrees with the above conclusions and the results in Ref.~\cite{dori,davidson,129,130}, which concludes that $\tan\beta$ is not a meaningful parameter in 2HDM. This is also the motivation for us to fully study the Baryogenesis scenario in 2HDM-$S_3$.

The main results of Ref.\cite{pal} are outlined as follows. The ratio parameter $\tan\beta$ does not cause the dynamics of phase transition and the effective potential should be written as two components, each corresponding to a VeV. However, $\tan\beta$ determines the value range of additional particles. The $S_3$ symmetry acts as a "boundary" between the two electroweak phase transitions.

To complete the Baryogenesis scenario in 2HDM-$S_3$ and continue the results of Ref.\cite{pal}, this paper presents the calculations of sphaleron and gravitational wave. Sphaleron is a process associated with the $B$ violation, so calculating and testing its rules to more firmly confirm the solution of baryon asymmetry problem in this model. Gravitational waves (GW) produced in the EWPT are almost a accompanying problem that can be connected to experimental data of current detectors.

The sphaleron energy functional is currently calculated in two different ways (the ansatz with scale-free parameters \cite{10} and the smooth one \cite{11,46}) but both require numerical solutions, which we summarize in Ref.\cite{pkll}. Similarly, the key parameters and calculation of gravitational wave energy density are also summarized in Ref.\cite{ptl}.

In addition, the effect of $S_3$ symmetry should be further evaluated even though dynamically the one has no impact. The difference among sphaleron energies and gravitational wave energy densities at the two phase transitions are considered. These bandgaps are sketched as phenomenologically the effects of $S_3$ symmetry.

The paper has the following structure. Except for the Introduction (Sec.~\ref{secInt}) and the Conclusion and Outlooks (Sec.~\ref{vi}), Appendix~\ref{a} and Sec.~\ref{iii} discusses the control of first order electroweak phase transition problem. This is a summary of the results in Ref.\cite{pal} and an analysis of the phase transition strength in values of $a$ to further confirm the independence of $S$ according to $a$. More importantly, we show how to control an first order electroweak phase transition. In Sec.~\ref{iv} we derive the sphaleron energy functional and the gravitational wave energy density which undergoes important parameters. Then we solve them and compare the results with existing data or evaluate them.

\section{Controling EWPT in The 2HDM with $S_3$ symmetry}\label{iii}

The search for a first order electroweak phase transition is essential, it outlines a possible path to the matter-antimatter asymmetry problem. This has been done and evaluated in Ref.\cite{pal}. In this section, the conditions for a first order EWPT will be comprehensively evaluated.
	
\subsection{The Higgs potential}\label{iiia}
	
The generic scalar potential of 2HDM-$S_3$ \cite{1601,melic} in the complex representation, has a simpler form that contains 5 real parameters:  
\begin{align}	
	V_\text{Higgs}^{\text{2HDM} \otimes S_3}\equiv \, V(\phi_1, \phi_2)=&\mu_1^2 (\phi_2\phi_2^\dagger + \phi_1\phi_1^\dagger) + \fr{1}{2} l_1 (\phi_2\phi_2^\dagger + \phi_1\phi_1^\dagger)^2 + \fr{1}{2} l_2 (\phi_2\phi_2^\dagger - \phi_1\phi_1^\dagger)^2 \crn
	&\qquad+ l_3 (\phi_1^{\dagger} .  \phi_2) (\phi_2^{\dagger} .  \phi_1) - \mu_2^2 \left( \phi_1^{\dagger} . \phi_2 + \phi_2^{\dagger} .  \phi_1  \right)+H.C.\label{1}
\end{align}

The above potential is obtained from adding $S_3$ symmetry and the $S_3$ soft violating potential to the initial position of 2HDM. The process of calculation and reduction to the above potential is the resetting of scalar fields and parameters presented in Refs.\cite{1601,melic}. Eq.\eqref{1} has mixed components between $\phi_1$ and $\phi_2$ but is much simpler than the original 2HDM potential.

The fields $\phi_1$ and $\phi_2$ have the following VEVs:
\begin{align}
	\label{VEV}
	\langle \phi_1 \rangle_0 =\fr 1{ \sqrt{2}}
	\begin{pmatrix}
		0\\
		v_1
	\end{pmatrix}
	,
	\langle \phi_2 \rangle_0 = \fr 1{ \sqrt{2}}
	\begin{pmatrix}
		0\\
		v_2
	\end{pmatrix}.
\end{align}

For a model with more than one VEV, there is usually a scaling parameter among the VEVs that are included to investigate phenomenological problems. In the model of interest, a very familiar parameter is introduced manually,
\begin{align}
	\tan \beta =\fr{s_\beta}{c_\beta}= \dfrac{v_2}{v_1}\,,
\end{align}
where $ s_\beta\equiv \sin\beta, \, c_\beta \equiv\cos\beta$. $\beta$ is the mixing angle between the charged Higgs and pseudoscalar particle. The value of $\tan\beta$ is summarized in the introduction secction.

When calculating the EWPT problem, the Higgs and gauge boson sector have the strongest contributions. After diagonalizing Eq.\eqref{1} and the gauge field sector \cite{pal}, the masses of Higgs fields are like Eq.\eqref{blkl}.

\begin{align}
	\begin{split}
		&m_{H^\pm}^2 =  -l_2 v^2, m_{A}^2 = - \dfrac{1}{2} (2l_2 - l_3) v^2 ,\\
		& m_{H}^2 = \dfrac{v^2}{4}
		\bigg[ 2 l_1 + l_3  + \sqrt{ 16 l_1 l_2 (c_\beta^2 - s_\beta^2)^2 - 8 l_2 l_3  + l_3^2 - 4 l_1 l_3 (c_\beta^4 - 6 c_\beta^2 s_\beta^2 + s_\beta^4) } \bigg]=\fr{f_H}{4}.v^2,\label{blkl}\\
		& m_{h}^2 = \dfrac{v^2}{4}
		\bigg[ 2 l_1 + l_3  - \sqrt{ 16 l_1 l_2 (c_\beta^2 - s_\beta^2)^2 - 8 l_2 l_3  + l_3^2 - 4 l_1 l_3 (c_\beta^4 - 6 c_\beta^2 s_\beta^2 + s_\beta^4) } \bigg]=\fr{f_h}{4}.v^2\,.
	\end{split}
\end{align}

The masses of $H^{\pm}$ and $A$ have no mixing between the two VEVs, this is not possible in the 2HDM model. Similarly, the gauge particles also do not have this mixing. However, we cannot eliminate this mixing for $H$ and $h$ either, so we are forced to approximate $f_H$ and $f_h$. 

The role of the $S_3$ symmetry is briefly summarized in the introduction. But we can see further that this symmetry is to make the squares of the masses of the $H^{\pm}$ and $A$ particles have no mixing component between the VEVs. In addition, this symmetry reduces the number of free parameters in the Higgs potential. Therefore, the $S_3$ symmetry has promoted the separation of the two electroweak phase transitions. This has been discussed in Ref.\cite{pal}.

The mass spectra of them can be derived by expanding the kinetic energy component of the higgs field. The mass spectra of all particles that contribute significantly to the EWPT process in Table \ref{mass1}.	
		\begin{table}[htp]
		\begin{tabular}{ |c|c|c|c|c|c| }
			\hline\hline
			Particles & $m(v_0)$ [GeV]&$m^2(v_1,v_2)$ & $m^2(v_1)$ & $m^2(v_2)$ & $n$ \\
			\hline\hline
			$W^\pm$ &$80.442$ &$\fr{g^2v^2}{4}$ &$\fr{g^2v_1^2}{4}$ &$\fr{g^2v_2^2}{4}$  & $6 $\\
			
			$Z$ &$91.18$ &$(g^2+ g'^2)\fr{v^2}{4}$ & $(g^2+ g'^2)\fr{v_1^2}{4}$ & $(g^2+ g'^2)\fr{v_2^2}{4}$ & $3$ \\
			
			$h$ &$125$ &$ \fr{1}{4} f_h v^2$ &$ \fr{1}{4} f_h v_1^2 $& $\fr{1}{4} f_h v_2^2$ & $1$ \\
			
			$H$ &$-$ &$ \fr{1}{4} f_H v^2$ &$ \fr{1}{4} f_H v_1^2 $& $\fr{1}{4} f_H v_2^2$ & $1$ \\
			
			$A$ &$-$ &$ - \dfrac{1}{2} (2l_2 - l_3) v^2  $ & $ - \dfrac{1}{2} (2l_2 - l_3) v_1^2  $ & $ - \dfrac{1}{2} (2l_2 - l_3) v_2^2  $ & $1$ \\
			
			$H^\pm$ &$-$& $ - l_2 v^2$ & $- l_2 v_1^2$ & $ - l_2 v_2^2 $ & $2$ \\
			
			$t$ &$173.1$ &$f_t^2 v^2$ &$f_t^2 v_1^2$ & $f_t^2 v_2^2$ & $-12$\\
			\hline\hline
		\end{tabular}	
		\centering
		\caption{Squared mass of the gauge bosons and scalar bosons in 2HDM-$S_3$; whereas mass of the $W^\pm$, $Z$ and $t$ is the same as the one in SM;  $v^2 = v_1^2 + v_2^2$; $v_0=246$ GeV.}	
		\label{mass1}
	\end{table}
	
In this model $v^2=v^2_2+v^2_1$, we put $v^2=a.v^2_2$. Therefore, the masses of particles in terms of $a$ are given in Table \ref{mass12}.
	
	\begin{table}[htp]
		\centering
		\begin{tabular}{|c|c|c|c| }
			\hline\hline
			Particles & $m^2(v_1,v_2)$ & $m^2(v_2)$ & $m^2(v_1)$  \\
			\hline\hline
			$m^2_{W^\pm}$ & $\fr{g^2v^2}{4}$ &$m^2_{W^\pm}/a$ &$m^2_{W^\pm}(v_2).(a-1)$\\
			
			$m^2_Z$ & $(g^2+ g'^2)\fr{v^2}{4}$ & $m^2_Z/a$ & $m^2_{Z}(v_2).(a-1)$  \\
			
			$m^2_h$ & $ \fr{1}{4} f_h v^2$ & $m^2_h/a$ & $m^2_{h}(v_2).(a-1)$  \\
			
			$m^2_H$ & $ \fr{1}{4} f_H v^2$ & $m^2_H/a$ & $m^2_{H}(v_2).(a-1)$ \\
			
			$m^2_A$ & $ - \dfrac{1}{2} (2l_2 - l_3) v^2  $ & $m^2_A/a$ & $m^2_{A}(v_2).(a-1)$   \\
			
			$m^2_{H^\pm}$ & $ - l_2 v^2$  & $m^2_{H^\pm}/a$ & $m^2_{H^{\pm}}(v_2).(a-1)$  \\
			
			$m_t^2$ & $f_t^2 v^2$ & $m^2_t/a$ & $m^2_{t}(v_2).(a-1)$ \\
			\hline\hline
		\end{tabular}
		\caption{Squared mass of the gauge and scalar bosons in 2HDM-$S_3$.}
		\label{mass12}
	\end{table}
\subsection{The temperature effective Higgs potential and controlling EWPT}

We see that there is a problem in examining the phase transition. Why is $v_1$ different from $v_2$ (meaning the vacuum breaking process of the two Higgs fields is different) that we examine them together? We examine the electroweak phase transition with two stages, because of the following two reasons: the first, as the above sections pointed out, in 2HDM-$S_3$, the masses of particles ($H$ and $h$) can be changed such that there are no mixing terms between $v_1$ and $v_2$. This approximation is similar to the way of examining a phase transition where $v_1$ and $v_2$ are both brought to the variable $v$. The second, we do not ignore the mixing of VEVs in the mass components, we only approximate them.

In 2HDM-$S_3$, most of the squares of the masses of the particles are split into two separate parts (the first part depends on $v^2_1$, the second part depends on $v^2_2$), except for the two particles $H$ and $h$ as seen in Eq.(\ref{blkl}). Therefore, the effective potential can be split into two separate components as in Eq.(\ref{5}), which is detailed in Ref.\cite{pal}, 
	
\be
V_{eff}^{S_3}\approx  V_{eff}^{S_3}(v_1)+V_{eff}^{S_3}(v_2).\label{5}
\ee

We assume $v_1<v_2$, so that the electroweak phase transition corresponding to $v_2$ occurs first, followed by the second phase corresponding to $v_1$. This assumption is actually the same as the opposite case $v_2<v_1$, in which case the electroweak phase transition corresponding to $v_1$ will occur before the phase corresponding to $v_2$ or we just need to change $v_1$ to $v_2$ in the initial case and vice versa. Therefore the assumption $v_1<v_2$ is also sufficient to investigate the electroweak phase transition. If $v_1=v_2$, both phase transitions occur simultaneously.

Also combined with the results of $\tan\beta$ given in the introduction, we see that $\tan\beta=v_2/v_1>1$ so it is reasonable to assume $v_2>v_1$.

Since we assume $v_1<v_2$ and $v_2<v, v^2=v_1^2+v^2_2$. We deduce that $a=v^2/v_2^2=1+v_1^2/v^2_2$ so $1<a<2$. However, for the cases where $v_1>v_2$ we just need to swap $v_1$ to $v_2$ and vice versa.

The mixing components between $v_1$ and $v_2$ in the masses of the $H$ and $h$ particles will be of the form $const.v_1.v_2$. If we express $v_1$ with respect to $v_2$ in terms of $a$, then these mixing components can be approximated as $const.v^2_2$ or $const.v^2_1$. These mixing components are therefore pushed into one of the two phase transitions. We therefore say that the mixing components between $v_1$ and $v_2$ have been absorbed into $V^{S_3}_{eff}(v_1)$ and $V^{S_3}_{eff}(v_2)$ when using $a$, as shown in Table \ref{mass12}. Hence in Eq.\ref{5}, although the effective potentials are separated, the mixing components are still present in them. In addition, although the electroweak phase transition is investigated with two phases, these phases may interact with each other through the mixing components in the mass components of the particles. This will be investigated further in this paper.

With the above formula for the potential, the phase transition strength does not depend on $a$. This has been proven in Ref.\cite{pal}. The high-temperature effective potential when considering daisy loops is rewritten as follows,
\begin{eqnarray}
	V_{eff}(\phi, T)&=&\lambda(T)\phi^4+D(T^2-T_0^2)\phi^2-\frac{T}{12\pi}\sum_{h,W,Z,H,H^\pm,A}g_i\left[\frac{m_i^2\phi^2}{v^2_{b0}}+\Pi_i(T)\right]^{\frac{3}{2}}.
\end{eqnarray}
where, $b=1,2$ then $v_{10}, v_{20}$ are $v_1, v_2$ at $T=0K$ respectively and
\begin{align}
	&\lambda(T)&=&\ \frac{m_h^2+m_H^2}{8v^2_{b0}}+\frac{1}{64\pi^2v^4_{b0}}\sum_{h,W,Z,H,H^\pm,A,t}g_im_i^4 \ln{\left(\frac{A_iT^2}{m_i^2}\right)}
	,\label{lambdae}\\
	&D&=&\ \frac{m_h^2+6m_W^2+3m_Z^2+m_H^2+2m_{H^\pm}^2+m_A^2+6m_t^2}{24v^2_{b0}}
	,\notag\\
	&T_0^2&=&\ \frac{1}{D}\left[\frac{m_h^2+m_H^2}{4}-\sum_{h,W,Z,H,H^\pm,A,t} \frac{g_im_i^4}{32\pi^2 v^2_{b0}}\right]
	,\notag\\
	&\Pi_W(T)&=&\ \frac{22}{3}\frac{m_W^2}{v^2_{b0}}T^2,\quad \Pi_Z(T)=\frac{22}{3}\frac{(m_Z^2-m_W^2)}{v^2_{b0}}T^2
	,\notag\\
	&\Pi_h(T)&=&\ \frac{m_h^2+2m_W^2+m_Z^2+2m_t^2}{4v^2_{b0}}T^2
	,\notag\\
	&\Pi_A(T)&\sim&\frac{m_A^2}{v^2_{b0}}T^2,\quad \Pi_{H^{\pm}}(T)\sim\frac{m_{H^{\pm}}^2}{v^2_{b0}}T^2,\quad \Pi_H(T)\sim\frac{m_H^2}{v^2_{b0}}T^2.\notag
	\end{align}

In Eq.\ref{lambdae}, $\ln A_i=3.907$ or $\ln A_i=1.135$ for bosons or fermions. $\Pi_W(T)$, $\Pi_Z(T)$, $\Pi_A(T)$, $\Pi_{H^{\pm}}(T)$, $\Pi_H(T)$ are functions that characterize the daisy loop contributions. The daisy loops of charged Higgs $(H^{\pm})$, neutral Higgs $(H)$ and $A$ are omitted because the transition temperatures less than 200 GeV, the contribution of these loops is negligible. This has been clearly demonstrated in the articles \cite{pal, phong2022}. Also with another method of calculating the thermal phase transition, in Ref.\cite{twoloop} without daisy loop components.

In addition, the calculation of the effective potential with the daisy loops of the heavy Higgs particles in the temperature effective potential is necessary for completeness of the problem. However, we need to see that the main driving force of these particles is the trigger for the EWPT process. Therefore, adding these loops may complicate the main calculation because their contributions are negligible in the context of the EWPT calculation.

From many conclusions, we propose a way to investigate or control the Higgs potential to have a first order phase transition or $S=\frac{v_c}{T_C}>1$ ($v_c$ and $T_C$ are VEV and critical temperature respectively): the first is to build the high temperature effective potential. The second is to select the investigation variable, usually the masses of new particles. Finally, survey in the absence of daisy loops. Specify the range of values for variables to have $S>1$. If the masses of new particles are greater than 200 GeV, there are no need to use additional daisy loops.

\begin{figure}[H]
	\centering
	\includegraphics[width=0.8\textwidth]{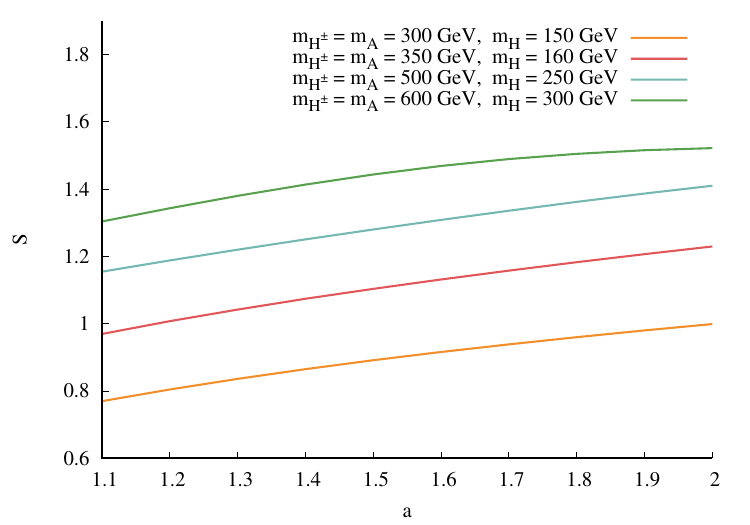}
	\caption{The strength of EWPT is a function of $a$. The masses of particles are in GeV. $m_{H^\pm}, m_A$ and $m_H$ in the figure are just $m_{H^\pm}(v_2), m_A(v_2)$ and $m_H(v_2)$.}\label{fig:1}
\end{figure}

In Fig. \ref{fig:1}, the strength $S$ almost does not change with $a$, for different mass cases. This is another way to confirm the results of previous article (in Ref.\cite{pal}). Also in Fig.\ref{fig:1}, the lines are not absolutely parallel to the $a$ axis (the slopes are very small). This means that $S$ still depends on $a$, because the $\Pi$ functions which are the contributions of daisy loops, still depend on $a$ but are negligible. Through such analysis, also following many conclusions from other authors, it is once again fully confirmed that the contribution of daisy loops in the vicinity of critical temperature $(T_C\sim 100-150$ GeV) is negligible.

With different mass pairs, the values of phase transition strength do not vary too much. These masses are arbitrary but falls within the range of values drawn from other studies. Not any mass pair of two particles gives a first order EWPT. The three parameters including $a$ and the two mass parameters are bound together. This is one of the conditions that can be used as a reference for the mass range of new particles. Thus the search for a first order electroweak phase transition depends on the masses of particles and even $a$ (when considering daisy loops). 

As in Ref.\cite{pal}, the value of $a$ is in the range 1 to 2. $a$ (or $\tan\beta$) is the parameter that is entered by hand, in the electroweak phase transition problem, although it has no role in influencing the strength of phase transition, but physically, it characterizes the distance between two phase transitions.  

There are two ways to control a first order phase transition: the first, choose the masses of the particles (which must also be large enough), then control $a$ to find $S>1$ as in this article (although the dependence on $a$ is quite small when taking into account daisy loops); the second, which is the way done in Ref. \cite{pal}. For models with multiple phase transitions, we should do both of these to cross-check the results.

\begin{figure}[H]
	\centering
	\subfloat[\centering  $a=1.1$]{{\includegraphics[width=7.5cm]{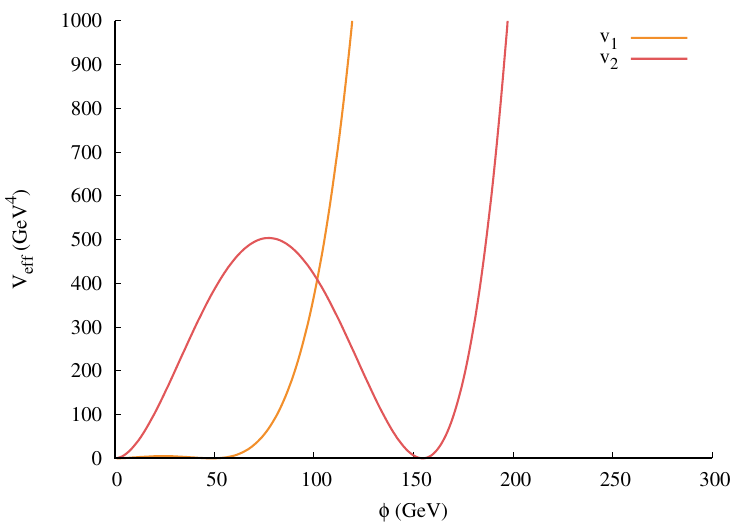} }}%
	\subfloat[\centering $a=1.3$]{{\includegraphics[width=7.5cm]{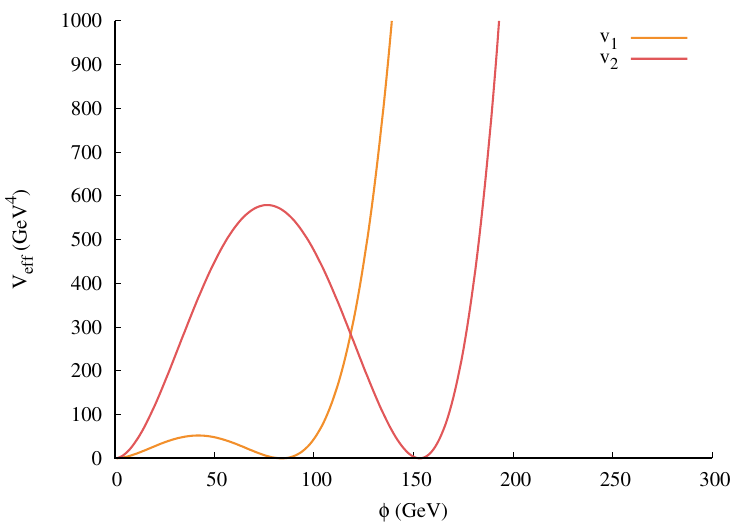} }}%
	
	\subfloat[\centering $a=1.5$]{{\includegraphics[width=7.5cm]{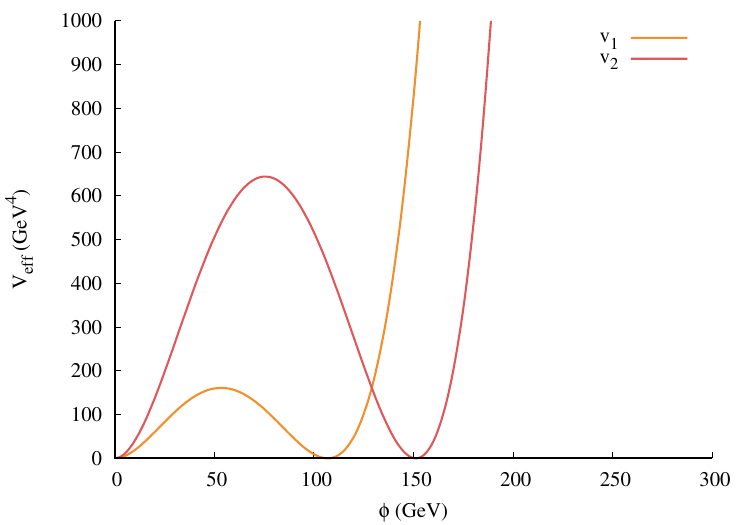} }}%
	\subfloat[\centering $a=1.9$ ]{{\includegraphics[width=7.5cm]{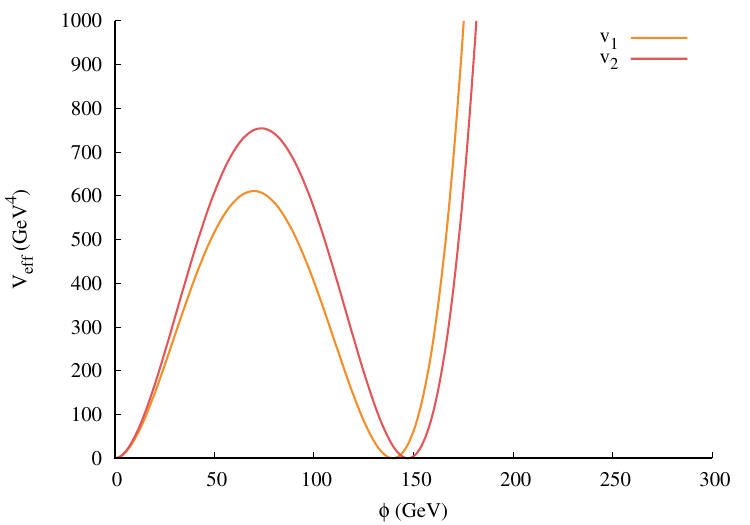}}}
	\caption{The effective potential $V_{eff}$ at $T=T_C$. $m_{H^{\pm}}(v_2)=m_A(v_2)=350$ GeV, $m_H(v_2)=160$ GeV. The red line corresponds to $V_{eff}(v_2)$. The orange line corresponds to $V_{eff}(v_1)$.}
	\label{fig:2a}
\end{figure}

According to Fig. \ref{fig:2a}, the two potential functions corresponding to $v_2$ and $v_1$ get closer to each other as $a$ increases. With the uniform convergence of second minima (i.e. always less than $246$ GeV), the two effective potentials are always similar to each other. Although Fig.\ref{fig:2a} only corresponds to one mass pair, it is quite general, other mass pairs are presented in Appendix \ref{a}.

Replacing the investigation in terms of $\tan\beta$ with $a$ actually gives an advantage in sketching the evolution of two components in the effective potential. Mathematically, the value domain of $\tan\beta$ is an open one (i.e. there is no upper bound). But when examined in terms of $a$, the value domain of $a$ is closed, it lies between 1 and 2. This change is like a mapping of $\tan\beta$ into $a$. In summary, finding a first order electroweak phase transition with an effective potential pair in 2HDM-$S_3$ is mathematically and physically feasible.

\begin{table}[H]
\centering
\resizebox{\textwidth}{!}{\begin{tabular}{||c||c|c|c|c|c|c|c||}\hline \hline
		a&$v_c$ [GeV]&$T_C$ [GeV]&$S$&$E_{sph,b}^T$ [GeV]&$\alpha$&$\betah$&$\Omega h^2 (f_{peak})\times 10^{-14}$ \\
		\hline
		\hline
		\multirow{2}{*}{1.1}
		&$v_2$=154.618	&	159.443	&	0.969736	&	7841.47	&	0.017436	&	22.1311	&	1.82991\\
		&$v_1$=48.8945	&	50.4204	&	0.969735	&	2479.69	&	0.0174359	&	22.1311	&	1.8299\\
		\hline
		\multirow{2}{*}{1.2}
		&$v_2$=153.951	&	152.784	&	1.00764	&	7588.49	&	0.019334	&	22.3506	&	2.70965\\
		&$v_1$=68.8489	&	68.3271	&	1.00764	&	3393.67	&	0.019334	&	22.3506	&	2.70965\\
		\hline
		\multirow{2}{*}{1.3}
		&$v_2$=153.12	&	146.913	&	1.04225	&	7365.48	&	0.0211846	&	22.5607	&	3.82934\\
		&$v_1$=83.8672	&	80.4677	&	1.04225	&	4034.24	&	0.0211846	&	22.5607	&	3.82934\\
		\hline
		\multirow{2}{*}{1.4}
		&$v_2$=152.188	&	141.682	&	1.07415	&	7166.8	&	0.02299	&	22.7626	&	5.21179\\
		&$v_1$=96.2521	&	89.6077	&	1.07415	&	4532.68	&	0.02299	&	22.7626	&	5.21179\\
		\hline
		\multirow{2}{*}{1.5}
		&$v_2$=151.2	&	136.979	&	1.10382	&	6988.23	&	0.0247537	&	22.9576	&	6.87871\\
		&$v_1$=106.914	&	96.8587	&	1.10382	&	4941.43	&	0.0247537	&	22.9576	&	6.87871\\
		\hline
		\multirow{2}{*}{1.6}
		&$v_2$=150.187	&	132.717	&	1.13164	&	6826.53	&	0.0264803	&	23.1465	&	8.85171\\
		&$v_1$=116.335	&	102.802	&	1.13164	&	5287.81	&	0.0264803	&	23.1465	&	8.85171\\
		\hline
		\multirow{2}{*}{1.7}
		&$v_2$=149.173	&	128.83	&	1.15791	&	6679.16	&	0.0281746	&	23.3302	&	11.1534\\
		&$v_1$=124.807	&	107.787	&	1.15791	&	5588.19	&	0.0281746	&	23.3302	&	11.1534\\
		\hline
		\multirow{2}{*}{1.8}
		&$v_2$=148.175	&	125.264	&	1.1829	&	6544.11	&	0.0298422	&	23.5091	&	13.8084\\
		&$v_1$=132.531	&	112.039	&	1.1829	&	5853.23	&	0.0298422	&	23.5091	&	13.8084\\
		\hline
		\multirow{2}{*}{1.9}
		&$v_2$=147.203	&	121.977	&	1.20681	&	6419.72	&	0.0314887	&	23.6838	&	16.8449\\
		&$v_1$=139.649	&	115.717	&	1.20681	&	6090.28	&	0.0314887	&	23.6838	&	16.8449\\
		\hline
		\multirow{2}{*}{2}
		&$v_2$=146.268	&	118.934	&	1.22982	&	6304.68	&	0.03312	&	23.8544	&	20.2955\\
		&$v_1$=146.268	&	118.934	&	1.22982	&	6304.68	&	0.03312	&	23.8544	&	20.2955\\
		\hline
		\hline
\end{tabular}}
	\caption{Results in case $m_{H^{\pm}}(v_2)=m_A(v_2)=350$ GeV, $m_H(v_2)=160$ GeV.}\label{omega}
\end{table}

In Fig.\ref{fig:2a}, the non-zero minimum of effective potential corresponding to $v_2$ is always larger than that of $v_1$. The yellow lines in Fig.\ref{fig:2a} approach the red lines as $a$ increases or the two phase transitions get closer together. When $a=2$, the two potential functions coincide, but then the electroweak phase transition is the simultaneous phase transition of two VEVs. This corresponds to the existence of two scalar Higgs fields in this model. It is interesting to note that these two scalar fields are partly involved in both phase transitions. No one scalar particle is involved in only one phase transition. The nature of this problem is that the higgs potential contains up to two Higgs doublets and is a general mixture of these two doublets. Furthermore, in Table \ref{omega}, for a given value of $a$, $S$ of the two EWPT stages are the same. This confirms the results in Ref.\cite{pal}. However, to find the maximum value of $S$ which is often not very precise, the method in Ref.\cite{pal} gives better results.

\section{Sphalerons and Gravitational waves in the 2HDM-$S_3$}\label{iv}

The 2HDM model with the $S_3$ symmetry shows a "twin bubble nucleation or double bubble nucleation". The first possible demonstration of this is that the masses of particles depends on the two VEVs. Two bubbles can form at the same time or one after the other. The following numerical analysis will clarify that.

\subsection{Sphaleron}
The sphaleron energy functional consists of three components: the contribution of gauge fields, Higgs kinetic energy and effective potential \cite{pkll,phong2022},
\be
E_{sph}=\int dx^3 \left[ \fr{1}{4}W_{ij}^e W_{ij}^e +(D_i \phi)^\dagger (D_i \phi)+ V_{eff}\right].
\label{e1}
\ee

The contributions of exotic Higgs bosons ($H^{\pm}$, $A$ and $H$) can be ignored. However, they still contribute to the effective potential components. The static field approximation (ie., $W^e_0=0$) has been assumed and sphaleron has a spherically symmetric form \cite{11}:
\be
\left\{ \begin{array}{ll}
	\phi(r)=\fr{v_0}{\sqrt{2}} h(r) i n_e \sigma^e \left(\begin{array}{cc} 0\\1\end{array} \right),\\
	W_i ^e (r)=\fr{2}{g}\ep ^{eij} n_j \fr{f(r)}{r},
\end{array}\right.
\label{ham}
\ee
where $n_i \equiv \fr{x^i}{r}$ and $r$ is the radial coordinate in the spherical coordinates. The index $e$ runs from 1 to 3. $\sigma^e$ are Pauli matrices. The functions $h(r), f(r)$ are profile functions, which characterize bubbles. 

From Eqs. (\ref{e1},\ref{ham}), according to Refs.\cite{pkll,phu5gw,phong2022}, the Sphaleron energy functional at temperature $T$ is written as follows
\begin{equation}
E_{sph}^{T}=\fr{4 \pi v_0}{g} \int_0^\infty d\xi \left[ 4\left(\fr{df}{\xi}\right)^2
+\fr{8 f^2(1-f)^2}{\xi^2}  + h^2(1-f)^2+\fr{\xi^2}{2}\left(\fr{dh}{\xi}\right)^2
 +\fr{\xi^2}{g^2 v^4_0}V_{eff}(h,T)\right],
\label{sphT}
\end{equation}
where $g^2=\fr{G_F 8 m^2_W}{\sqrt{2}}$; $G_F=1.166 \times 10^-{15} \, \textrm{GeV}^{-2}$; $m_W=80.39\, \textrm{GeV}$, $\xi\equiv gv_0r$.\\

To make the above function converge in the process of solving, the last term is rewritten as a difference. Thus specifically for each stage will be

\begin{align}
	E^T_{sph,b}=\frac{4\pi v_{b0}}{g}\int_0^\infty d\xi\ &\bigg[ 4\left( \frac{df_b}{d\xi} \right)^2+\frac{8}{\xi^2}f^2_b(1-f_b)^2+h^2_b(1-f_b)^2+\frac{\xi^2}{2}\left( \frac{dh_b}{d\xi} \right)^2\nonumber\\
	&+\frac{\xi^2}{g^2v_{b0}^4}(V_{eff}(v_{c,b}h_b,T_{c,b})-V_{eff}(v_{c,b},T_{c,b}))\bigg].\label{11}
\end{align}

Note that in Eq. (\ref{11}), the run index $b=1,2$, it is the phase transition of $v_{1,2}$, respectively.

Eq.(\ref{sphT}) may diverge. To handle the divergences, especially in the last term of Eq.(\ref{sphT}), i.e. $V_{eff}(h,T)$, we have to rewrite Eq.(\ref{sphT}) as Eq.(\ref{11}). We do so, because it involves the equations of motion (\ref{em1}), (\ref{em2}) and the boundary conditions (\ref{dk}).

Observing the form of Eq.(\ref{sphT}), as $f$ approaches 1, the first three terms in Eq.(\ref{sphT}) converge. Therefore, we deal with the divergence in the last term of Eq.(\ref{sphT}) by choosing boundary conditions for $h$ and finding temperature-dependent factors in $V_{eff}(h,T)$ to eliminate them. This process is detailed in Refs.\cite{pkll,phu5gw,phong2022}.

Taking the variation of Eq.(\ref{11}) in terms of $h_b$ and $f_b$, after minimizing the Sphaleron functional, the equations of motion are obtained,

	\begin{align}
		\frac{d^2f_b}{d\xi^2}&=\frac{2}{\xi^2}f_b(1-f_b)(1-2f_b)-\frac{1}{4}(1-f_b)h_b^2,\label{em1}\\
		\frac{d^2h_b}{d\xi^2}&=\frac{-2}{\xi}\frac{dh_b}{d\xi}+\frac{2}{\xi^2}h_b(1-f_b)^2+\frac{1}{g^2v_{b0}^4}\frac{\partial V_{eff}(v_{c,b}h_b,T_{c,b})}{\partial h_b}.\label{em2}  
	\end{align}

Here, $h_b(\xi)$ and $f_b(\xi)$ have boundary conditions,
\be
\left\{ \begin{array}{ll}
	h_b(\xi\to 0)=f_b(\xi\to 0)=0,\\
	h_b(\xi\to \infty)=f_b(\xi\to \infty)=1.
\end{array}\right.
\label{dk}
\ee

\begin{figure}[H]
	\centering
	\subfloat[\centering Profile functions with $a=1.1$]{{\includegraphics[width=8cm]{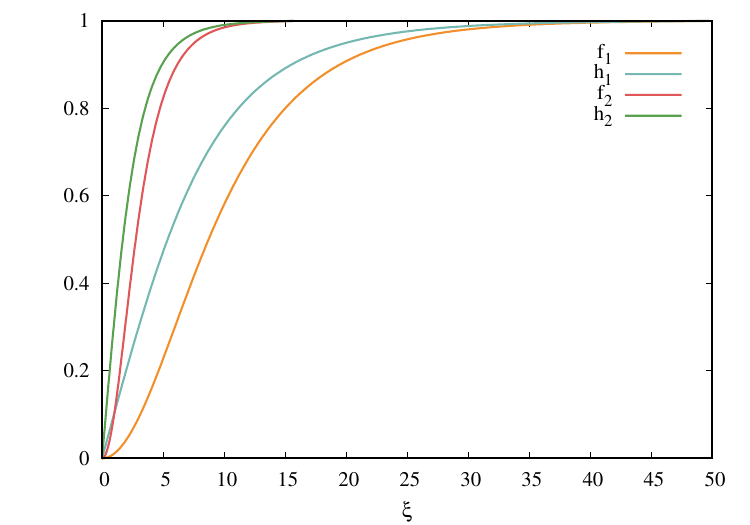} }}%
	\subfloat[\centering Profile functions with $a=1.3$]{{\includegraphics[width=8cm]{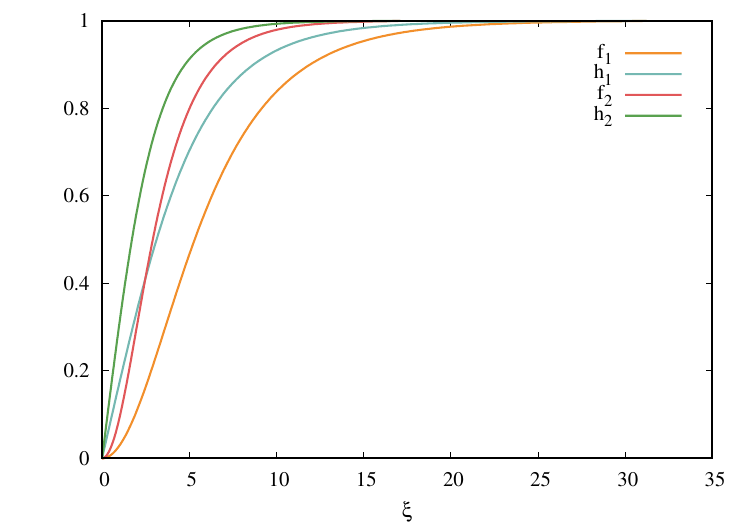} }}\\
	\subfloat[\centering Profile functions with $a=1.5$]{{\includegraphics[width=8cm]{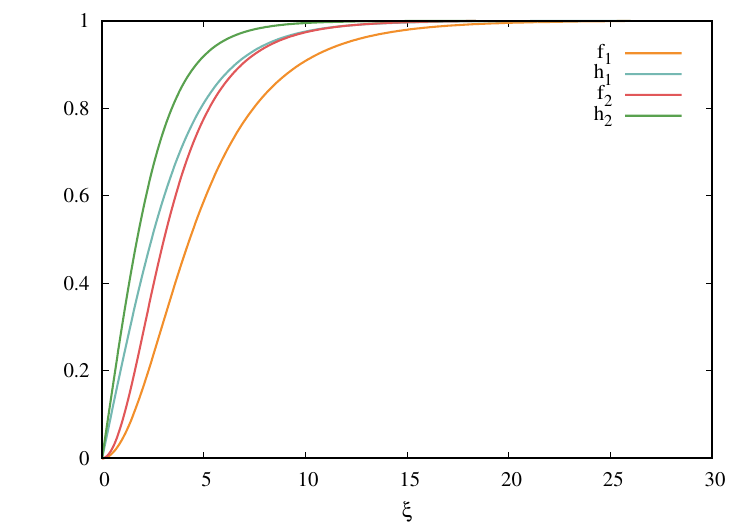} }}
	\subfloat[\centering Profile functions with $a=1.9$]{{\includegraphics[width=8cm]{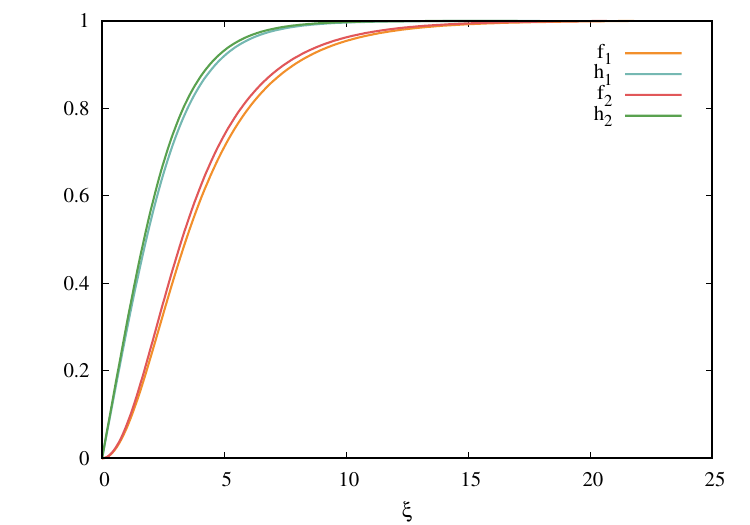} }}%
	\caption{$m_{H^{\pm}}(v_2)=m_A(v_2)=350$ GeV, $m_H(v_2)=160$ GeV.}\label{fig:3}
\end{figure}

The physical motivation for the boundary conditions: Profile functions represent the expansion process of bubble nucleations (BN). Inside the BN, the particles are in the mass phase, the particles are in the massless phase outside the bubbles. The bubbles gradually grow and touch each other, then all the particles are in the mass phase, so when the bubbles (represented by profile functions) grow to a certain value, all the particles are in the mass phase. They cannot grow forever, the BN cannot increase forever according to $\xi$ as Fig.\ref{fig:3}.

From Eqs. (\ref{em1},\ref{em2},\ref{dk}), the numerical solutions are obtained as Fig.\ref{fig:3}.

In Fig.\ref{fig:3}, the blue lines approach the green lines, the red lines go to the orange lines, as $a$ increases. The blue and green lines represent the bubbles of gauge fields, the orange and red lines represent the bubbles of Higgs fields, corresponding to the two phases.

When the phase transition occurs, $v_2$ is broken before $v_1$. The blue and red lines (corresponding to the bubbles associated with $v_2$) approach 1 faster or erlier than the green and orange lines (corresponding to the bubbles associated with $v_1$).

In other words, the process of forming bubble nucleations, to give the masses of particles, from massless one: the first they form a bubble nucleation corresponding to $v_2$ first, then form a bubble nucleation corresponding to $v_1$ so the initial particles form a pair of bubble nucleation. But these two nucleations can form at the same time when $a=2$, we can name this pair as a twin bubble nucleation.

And the sphaleron energy as the fifth column in Table \ref{omega}. According to Appendix \ref{a}, $a$ and the sphaleron energy both increases. The total energy value of these two phase transitions is in the range of 10 to 15 TeV.

\subsection{Gravitational waves}

The formulas to calculate the gravitational wave energy density (GWED) will be systematized as a setting for the calculation process. There are three processes in the production of GWs with a first-order EWPT: the bubble wall collisions (Coll), sound waves (SW) and Magnetohydrodynamic (MHD) turbulence \cite{71}. Usually they are added linearly to calculate the gravitational wave energy density \cite{71}:	
\be
h^2\Om_{GW}=h^2\Om_{Coll}+h^2\Om_{sw}+h^2\Om_{tur}.
\ee	

The GWED of SW depends on $\al $ and $\beta/H^*$ \cite{phu5gw, 71}:
\be
h^2\Om_{sw}(f) \simeq 2.65 \times 10^{-6} V_{gn}k^2_w \left[\fr{H_*}{\beta}\right] \left[\fr{\al }{1+\al }\right]^2\left[\fr{100}{g^*}\right]^{\fr{1}{3}}S_{sw}(f),
\label{ome}
\ee
where
\be
S_{sw}(f)=(f/f_{sw})^{3}\left(\frac{7}{4+3(f/f_{peak.sw})^2}\right)^{7/2},
\ee
and at the temperature $T_*$, the nucleation temperature \cite{71}, the peak frequency estimated as follows
\be
f_{peak.sw}=1.9\times 10^{-2} mHz \times \frac{1}{V_{gn}}\frac{\beta}{H^*}\frac{T_*}{100}\left(\frac{g_*}{100}\right)^{1/6}.
\ee

The bubble wall velocity is defined via $\al $ \cite{71}:
\be
V_{gn}= \fr{\fr{1}{\sqrt{3}}+\sqrt{\al ^2+2\fr{\al }{3}}}{1+\al }.
\label{al}
\ee

GWs from turbulence,
\be
h^2\Om_{tur}(f) \simeq 3.35\times 10^{-4} \left(\fr{H^*}{\beta}\right) V_{gn} \left(\fr{k_t\al }{1+ \al }\right)^{\fr{3}{2}}\left(\fr{100}{g^*}\right)^{1/3}S_{tur}(f),
\label{omet}
\ee
in which
\be
S_{tur}(f)=\frac{(f/f_{peak.tur})^{3}}{\left(1+(f/f_{peak.tur})\right)^{11/3}\left(1+8\pi f/h^*\right)} \label{tur},
\ee
and at $T_*$, Eq. (\ref{tur}) reads
\bea
h_{*} &= &16.5\times 10^{-3} mHz \times \frac{T_*}{100}\left(\frac{g_*}{100}\right)^{1/6}\,,\crn
f_{peak.tur} &=&2.7\times 10^{-2} mHz \times \frac{1}{V_{gn}}\frac{\beta}{H^*}\frac{T_*}{100}\left(\frac{g_*}{100}\right)^{1/6}.
\eea
One can estimate \cite{71} that
\be
k_t=(0.05-0.1)\times k_{w}; k_{w}= \al\left[0.73 +0.083\sqrt{\al}+\al\right]^{-1}, V_{gn} \sim 1.
\label{k}
\ee

BN growing in a plasma can attain a velocity ($V_{gn}\sim 1$). Here, the energy within the scalar field is insignificant (it only varies with the bubble's surface and not its volume), and the primary contributions to the signal are anticipated to come from the fluid's bulk movement. This may manifest as SW and MHD turbulence \cite{phu5gw}. In the case $V_{gn}\longrightarrow C$ ($C$ is the speed of light), we will also consider $\Omega_{coll}$ \cite{phu5gw}.

From the value of $\alpha$ in Table \ref{omega} and Appendix \ref{a}, $V_{gn}$ is around $0.75\sim 1$, that is, it can be compared to 1, or it has size 1. Therefore, the generation of gravitational waves has only two components: the sound wave and turbulence \cite{71, 73, 73a, 73b, 73c, 73d, 73e}. Thus, we need to calculate Eq.\eqref{ome} and Eq.\eqref{omet} or $\beta/H^*$ and $\alpha$.

The next, in Table \ref{Et}, the contribution of the Higgs field component to the total Sphaleron energy is on average about $45\%$ and by using a possible approximation \cite{phong2022}, because when calculating GW we only consider the effective potential and scalar field component \cite{71, phu5gw},	
\be
\fr{\beta}{H^*}\approx \left[T \fr{d\left(0.45\fr{E^T_{sph,b}}{T}\right)}{dT}\right]_{T=T^*},
\label{beta}
\ee

in which $\fr{\beta}{H^*}$ is a dimensionless quantity and is calculated using the sphaleron energy and the potential energy. $T^*$ is nucleation temperature. Additionally, in Ref. \cite{8gw}, at $T^*=T_C$, $H^*$ is the Hubble rate, is estimated as
\bea
H^*&=& 4.5 \times 10^{-22}\left[\fr{T_C}{MeV}\right]^2\,\textrm{ MeV} \left[\fr{g_*}{10.75}\right]^2;\,  g^*=106.75.
\label{hsao}
\eea

\begin{table}[H]
	\centering
	\resizebox{\textwidth}{!}{\begin{tabular}{|c|c|c|c|c|}
			\hline \hline
			\multirow{3}{*}{$a$}
			&\multicolumn{4}{c|}{$E^T_{Higgs,b}/E^T_{sph,b} (\%)$}\\
			\hline
			&$(m_{H^\pm}(v_2),m_H(v_2))$&$(m_{H^\pm}(v_2),m_H(v_2))$&$(m_{H^\pm}(v_2),m_H(v_2))$&$(m_{H^\pm}(v_2),m_H(v_2))$\\
			&$(600,300)[GeV]$&$(500,250)[GeV]$&$(350,160)[GeV]$&$(300,150)[GeV]$\\
			\hline
			\hline
			1.1&43.7192&44.1598&45.7986&46.6429	\\
			\hline
			1.2&43.6772&44.0209&45.5623&46.4284	\\
			\hline
			1.3&43.6652&43.9105&45.3493&46.2181	\\
			\hline
			1.4&43.6798&43.8245&45.1576&46.0172	\\
			\hline
			1.5&43.7177&43.7595&44.9851&45.8281	\\
			\hline
			1.6&43.7761&43.7129&44.8297&45.6517	\\
			\hline
			1.7&43.8521&43.6829&44.6897&45.488	\\
			\hline
			1.8&43.9425&43.6677&44.5634&45.3363	\\
			\hline
			1.9&44.0446&43.6661&44.4493&45.1959	\\
			\hline
			2&44.1555&43.677&44.3465&45.0659	\\
			\hline
			\hline
	\end{tabular}}
	\caption{The ratio between the Higgs component and the total sphaleron energy.}\label{Et}
\end{table}

In Table \ref{Et}, $E^T_{Higgs,b}$ is the components containing only the function $h_b(\xi)$ in Eq.\eqref{11}, in most cases $E^T_{Higgs,b}/E^T_{sph,b}$ decreases as $a$ increases. It ranges from $43.67\%$ to $46.64\%$. So the average of this ratio is around $45\%$. We therefore approximate it as Eq.\eqref{beta}. We also find that this ratio holds for most models because the contribution of the effective potential component to sphaleron is quite small. Indeed, this ratio in the THD model is exactly the same as the result in Ref. \cite{phong2022}.

The last parameter is $\al $  \cite{2gw, 2gwa,2gwb}, summarized as the below formulas:
\begin{align}
	\al &=\fr{\ep }{\rho_{rad}^*}, \rho_{rad}^*=g^* \pi^2 \fr{T^4_C}{30}=106.75\pi^2 \fr{T^4_C}{30},\label{all2}\\
	\ep &=\left(V_{eff}(v(T),T)-T \fr{d}{d T}V_{eff}(v(T),T)\right)_{T=T_C}.\label{all}	
\end{align}

$\alpha$ depends almost entirely on the effective potential. Therefore, in models, for a one-loop effective potential, $\alpha$ can be estimated under a value of $S$ and GW can be calculated at $T_C$. 

In conclusion, to calculate GW, we need to calculate $\beta/H^*$ and $\alpha$. $\beta/H^*$ is calculated through the sphaleron energy which is directly related to the effective potential. $\alpha$ is defined by Eqs. (\ref{all2},\ref{all}) which is related to the effective potential. Therefore, the effective potential contributes to GW.

\begin{figure}[H]
	\centering
	\subfloat[\centering  $a=1.1$]{{\includegraphics[width=8cm]{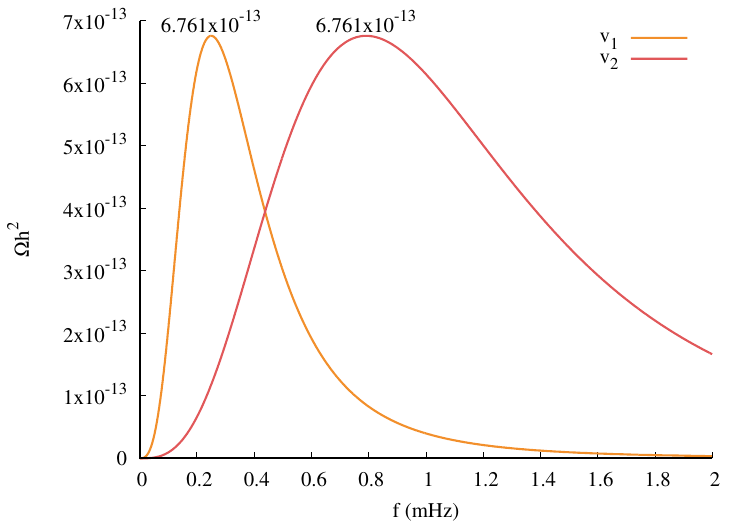}}}
	\subfloat[\centering  $a=1.3$]{{\includegraphics[width=8cm]{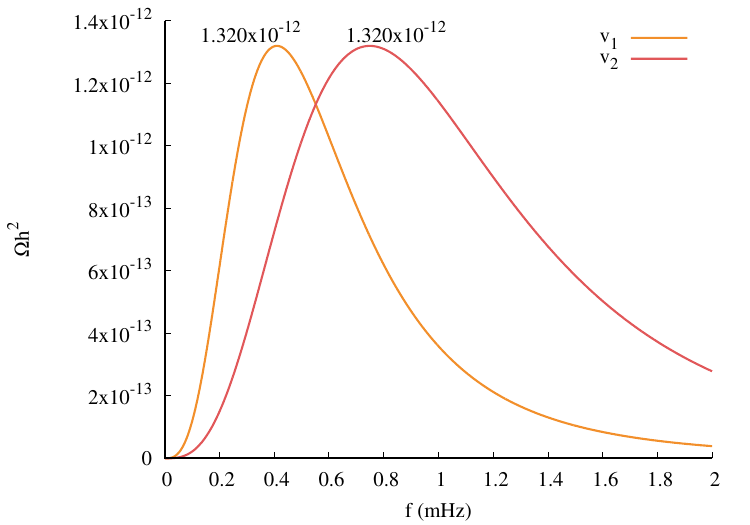} }}\\
	\subfloat[\centering  $a=1.5$]{{\includegraphics[width=8cm]{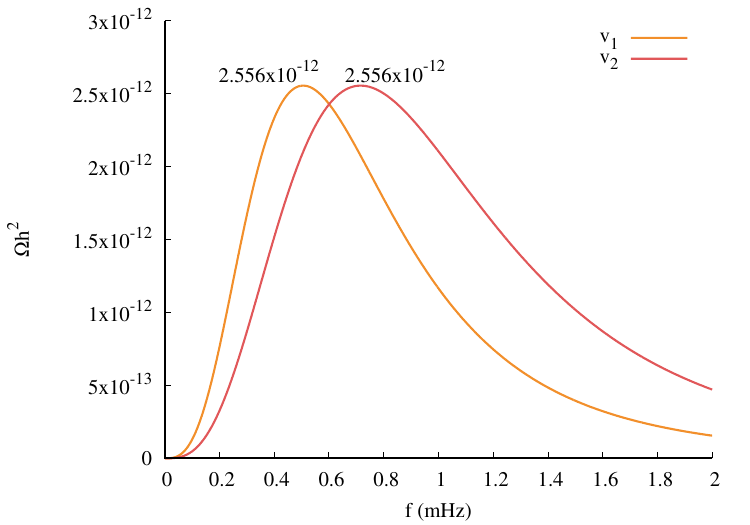} }}
	\subfloat[\centering $a=2$]{{\includegraphics[width=8cm]{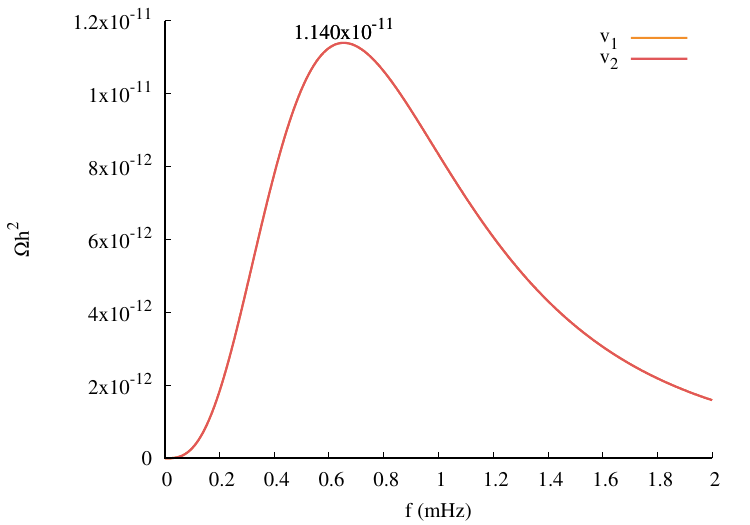} }}
	\caption{Gravitational wave energy density at $T^*=T_C$. The blue line corresponds to $v_2$, the purple line corresponds to $v_1$. $m_{H^{\pm}}(v_2)=m_A(v_2)=600$ GeV, $m_H(v_2)=300$ GeV.}\label{gwp}
\end{figure}

The two lines in each sub-figure in Fig.\ref{gwp} go closer to each other as $a$ increases. This is also consistent with the previous observations about the two EWPT stages.

From the numerical results in Table \ref{omega} and Appendix \ref{a}, the gravitational wave energy density of two phase transitions are equal at $f=f_{peak}$ at $T_C$. This is shown visually in Fig. \ref{gwp}. Because $a$ does not affect the phase transition strength of the two stages (i.s. the phase transition strength of the two stages is equal for one value of $a$). $\Omega h^2$ ranges widely, from $10^{-14}$ to $10^{-11}$. As $a$ increases, $\Omega h^2$ increases sharply.

From Appendix \ref{a}, as the larger the masses of the new particles get, the larger $S$ gets, i.e. they contribute more strongly to the effective potential, and the larger GW gets. In other words, the $H,h,H^\pm, A$ particles have made the effective potential have a large potential barrier in the electroweak phase transition, the more violent this process, the larger the gravitational wave generated.

\begin{figure}[H]
	\centering
	\subfloat[\centering  $\Omega h^2$ of the phase transition $v_1$ \cite{phu1gw}.]{{\includegraphics[width=8cm]{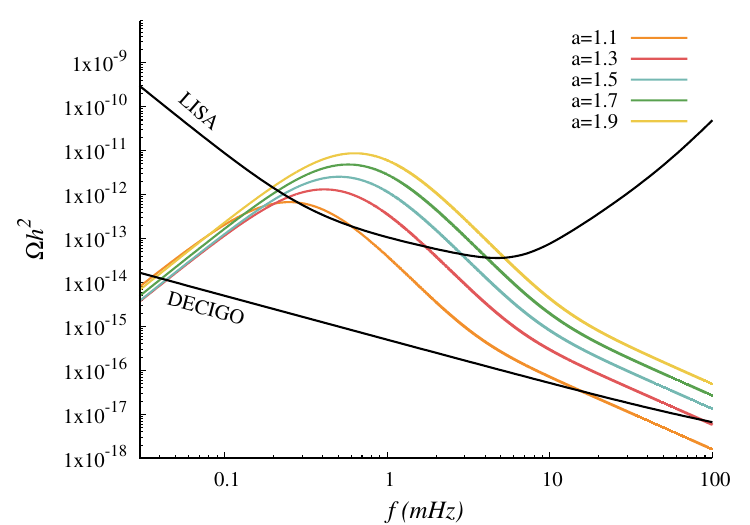} }}%
	\subfloat[\centering $\Omega h^2$ of the phase transition $v_2$ \cite{phu1gw}. ]{{\includegraphics[width=8cm]{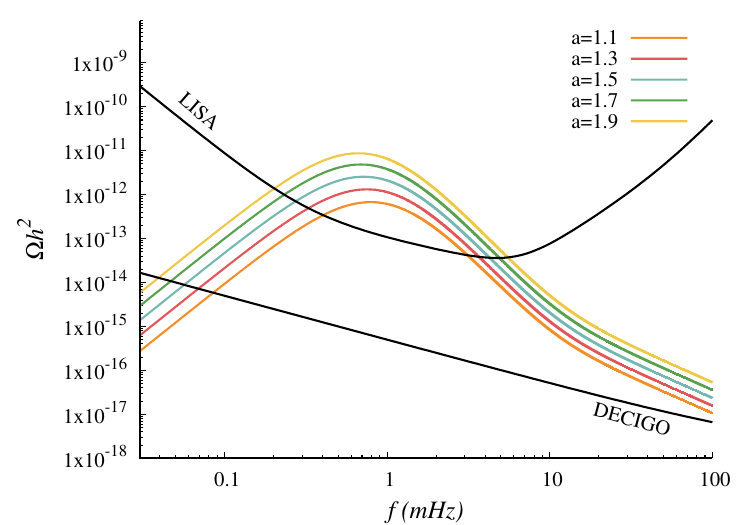}}}
	\caption{$\Omega h^2$ with $m_{H^{\pm}}(v_2)=m_A(v_2)=600$ GeV, $m_H(v_2)=300$ GeV.}
	\label{figgwe2hf}
\end{figure}

\begin{figure}[H]
	\centering
	\subfloat[\centering  Comparison with LISA \cite{phu1gw}.]{{\includegraphics[width=8cm]{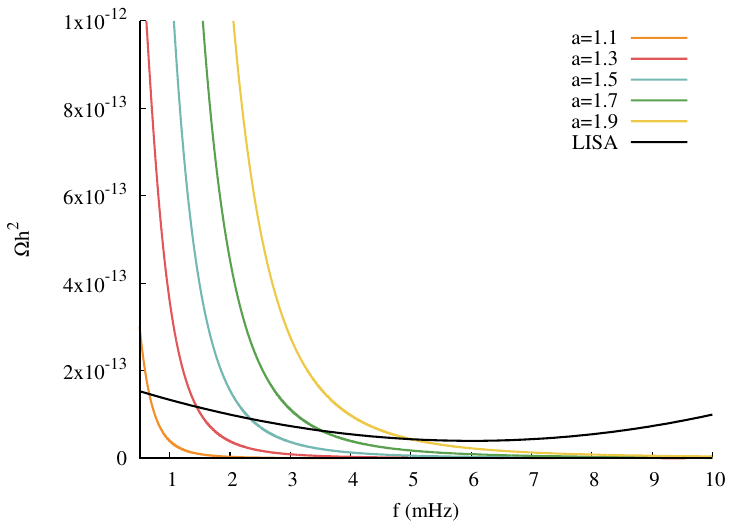} }}%
	\subfloat[\centering Comparison with DECIGO \cite{phu1gw}. ]{{\includegraphics[width=8cm]{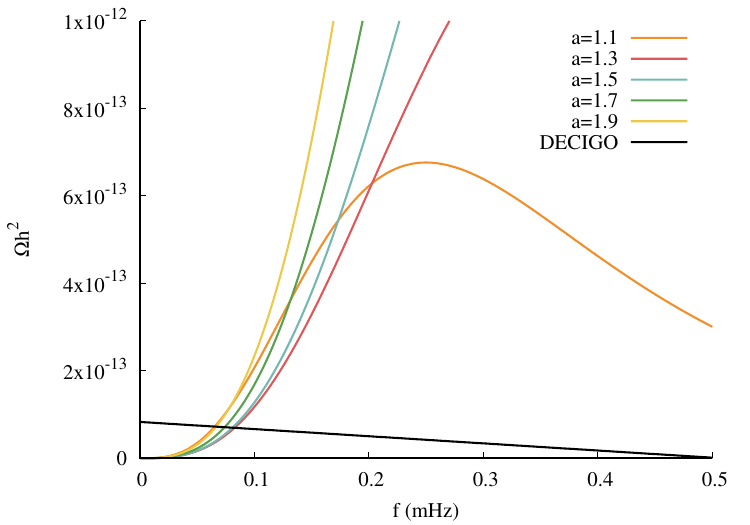}}}
	\caption{$\Omega h^2$ of the phase transition $v_1$. $m_{H^{\pm}}(v_2)=m_A(v_2)=600$ GeV, $m_H(v_2)=300$ GeV.}
	\label{figgwef}
\end{figure}

The gravitational wave detection frequency is solved in the range below a few mHz. At such low frequencies there is currently no experimental data. However, future detectors may be able to detect gravitational energy densities in the small frequency range as summarized in Table \ref{dulieu}. The LISA and DECIGO detectors can pick up the gravitational wave signal generated by the EWPT process in the 2HDM-$S_3$ model as shown in Figs. \ref{figgwe2hf}, \ref{figgwef}.

\begin{table}[!ht]
	\centering	
	\begin{tabular}{|c|c|c|c|}\hline\hline
		$f [mHz]$&$\Omega h^2$& Kind & Refs.\\ \hline	
		$0.02-0.12$& $2\times10^{-12}-10^{-10}$ & LISA& Refs.\cite{phu1gw,kudoh,thranel}\\
		$0.01-0.12$& $5\times 10^{-15}-8\times 10^{-14}$ & DECIGO & Refs.\cite{phu1gw,kudoh,thranel}\\
		$0-0.2$& Not yet & BBO & Refs.\cite{phu1gw,kudoh,thranel}\\
		\hline\hline
	\end{tabular}
	\caption{Sensitivity of proposed GW.}\label{dulieu}
\end{table}	

According to Fig. \ref{figgwef}, when $f\le 1$ mHz the gravitational wave results are consistent with DECIDO but when $f\sim$ a few mHz, LISA gives better results.

\section{CONCLUSION AND OUTLOOKS}\label{vi}

\begin{table}[!ht]
	\centering	
	\begin{tabular}{|c|c|c|c|c|c|c|c|} \hline\hline
		$a$&$v_{b0}$ [GeV]&$v_c$ [GeV]&$E_{b0}$ [GeV]&$E^{T_C}_{sph,b}$ [GeV]&$E^{T_C}_{sph,b}/v_c$&$E_{b0}/v_{b0}$& Difference \\ 
		\hline
		\multirow{2}{*}{1.3}	
		&$v_{20}=215.592$&$v_2$=153.12&7902.28&7365.48&48.10	&36.65	&\multirow{2}{*}{$24\%$}\\
		&$v_{10}=118.656$&$v_1$=83.8672&4302.69&4034.24&48.10	&36.26	&\\
		\hline
		\multirow{2}{*}{1.5}	
		&$v_{20}=200.96$&$v_2$=151.2&7311.87&6988.23&46.22	&36.38	&\multirow{2}{*}{$21\%$}\\
		&$v_{10}=142.026$&$v_1$=106.914&5172.53&4941.43&46.22	&36.42	&\\
		\hline
		\multirow{2}{*}{1.7}	
		&$v_{20}=188.665$&$v_2$=149.173&6826.74&6679.16&44.77	&36.18	&\multirow{2}{*}{$19\%$}\\
		&$v_{10}=157.979$&$v_1$=124.807&5707.53&5588.19&44.77	&36.13	&\\
		\hline\hline
	\end{tabular}
	\caption{$m_{H^{\pm}}(v_2)=m_A(v_2)=350$ GeV, $m_H(v_2)=160$ GeV.}\label{table:3}
\end{table}
	
According to the data tables, at the same value of $a$, the ratio $E^T_{sph,b}/v_c$ is always equal. By the linear scaling law of sphaleron energy \cite{scaling1, scaling2, scaling3}, $E^T_{sph,b}/v_b=E_{b0}/v_{b0}=const$ ($E_{b0}$ is the sphaleron energy at $T=0K$, Eq.\eqref{em2} calculated by replacing the effective potential at temperature $T$ with one at $0$K), the evolution (increase or decrease of sphaleron energy) in the two phase transitions is the same. However, $a$ and this progression increase proportionally. 

Additionally to test the scaling law (however, this was initially only an approximation to solve the Sphaleron energy problem \cite{scaling2, scaling3}). We estimate some cases like Table \ref{table:3}, $E^T_{sph,b}/v_b$ and $E_{b0}/v_{b0}$ differ by no more than $24\%$. Thus the results are not very compatible with the scaling law or this law is only true in a non-zero temperature range but is not true in the temperature range from 0 to $T_C$. For some other cases the calculation also breaks this law as in Ref.\cite{phascaling}. Because there are many triggers outside SM and energy scales other than SM. Furthermore, the solutions $h_b(\xi)$ and $f_b(\xi)$ in Eqs.~\eqref{em2} will approach $1$ very slowly at $0$, leading to $E_{b0}$ no longer holding the scaling law. Furthermore, since $v_{eff}(T)$ is only used for non-zero temperatures, that is, $V_{eff}(0)\neq \lim_{T\longrightarrow 0}v_{eff}(T)$.

The total sphaleron energy of two phase transitions is about $10-13$ TeV in this model. The value differs by about $1-5$ TeV from the SM. We also need to check the decoupling condition of sphaleron energy \cite{decoupling, decouplingb, decouplingc} as the following formula,
\begin{equation}\label{decoupling_equation}
	\frac{E_{sph,b}^T}{T}-7ln\left(\frac{v_b(T)}{T}\right)+ln\left(\frac{T}{100 GeV}\right)>(35.9-42.8).
\end{equation}

Based on the data tables in Appendix \ref{a}, the $\fr{E^T_{sph,b}}{T}$ ratio is always greater than $50$; $v_c$ is not much bigger than $T_C$ and almost $T_C$ are larger than $100$ GeV so $[-7\ln(v_c/T_C)+\ln(T/100)]$ has a negative value but not less than $-5$. Thus the sphaleron energy always satisfies the decoupling condition in Eq.\eqref{decoupling_equation}. Also, the nucleation temperature $T_N$ is not much larger than $T_C$, the supercooling process is small. In other words, the onset of phase transition is not so slow. However, the smaller $a$ is, the larger the gap between two phases is, the longer the EWPT is, another supercooling process in this model.

$\beta/H^*$ is actually quite difficult to calculate, it is estimated by us from the different contributions to the sphaleron energy. Based on numerical calculations, the contribution of gauge field component to this energy is $55\%$ so having an approximation as Eq.\eqref{beta}. In general an approximation, $\fr{\beta}{H^*}\approx T \fr{d}{dT}\left(\gamma.\fr{E^T_{sph,b}}{T}\right)$ can be choosed for other models; $\gamma$ depends on the contributions of components.

In the future, with new experimental data from the LHC, simply confirming the signals of particles beyond SM or sphaleron, it will indirectly confirm the Baryogenesis scenario.  

Thus the baryogenesis scenario in the 2HDM-$S_3$ model is fully calculated. One assertion is that this model has enough triggers to generate a first order electroweak phase transition, with sphaleron energies around the mid-range value of about 10 TeV which are checked by theoretical conditions and GW energies are compared with experimental possibilities. Additionally, the prospects for GW, which can be used for further reference, found in Ref.\cite{gwhite}.

Our future work follows the results, calculating CP violation, searching and connecting the strength of EWPT, sphalerons or gravitational waves with experimental data. Furthermore, we will extend these results in combination with some nonstandard cosmology scenarios and dark matter as in Refs.\cite{loc1, loc2}.

\section*{ACKNOWLEDGMENTS}
This research is funded by Vietnam National Foundation for Science and Technology Development (NAFOSTED) under grant number 103.01-2023.16.
\newpage
\appendix
\section{Table of data}\label{a}	
\begin{table}[H]
    \centering
\resizebox{0.95\textwidth}{!}{\begin{tabular}{||c||c|c|c|c|c|c|c||}\hline \hline
		$a$&$v_c$ [GeV]&$T_C$ [GeV]&$S$&$E$ [GeV]&$\alpha$&$\betah$&$\Omega h^2 (f_{peak})\times 10^{-15}$ \\
		\hline
		\hline
		\multirow{2}{*}{1.1}
		&$v_2$=128.327	&	166.643	&	0.770071	&	7621.73	&	0.0101885	&	20.5816	&	2.40066\\
		&$v_1$=40.5805	&	52.697	&	0.770071	&	2410.2	&	0.0101885	&	20.5816	&	2.40066\\
		\hline
		\multirow{2}{*}{1.2}
		&$v_2$=128.655	&	159.876	&	0.804718	&	7346.95	&	0.0114752	&	20.6793	&	3.81175\\
		&$v_1$=57.5363	&	71.4987	&	0.804717	&	3285.66	&	0.0114752	&	20.6793	&	3.81173\\
		\hline
		\multirow{2}{*}{1.3}
		&$v_2$=128.713	&	153.922	&	0.836219	&	7109.38	&	0.0127468	&	20.7847	&	5.727\\
		&$v_1$=70.4988	&	84.3067	&	0.836219	&	3893.97	&	0.0127468	&	20.7847	&	5.72701\\
		\hline
		\multirow{2}{*}{1.4}
		&$v_2$=128.568	&	148.628	&	0.86503	&	6901.35	&	0.0139997	&	20.8951	&	8.22445\\
		&$v_1$=81.3135	&	94.0008	&	0.86503	&	4364.8	&	0.0139997	&	20.8951	&	8.22446\\
		\hline
		\multirow{2}{*}{1.5}
		&$v_2$=128.272	&	143.878	&	0.891532	&	6717.06	&	0.0152319	&	21.0086	&	11.3772\\
		&$v_1$=90.7021	&	101.737	&	0.891532	&	4749.68	&	0.0152318	&	21.0086	&	11.3772\\
		\hline
		\multirow{2}{*}{1.6}
		&$v_2$=127.864	&	139.583	&	0.916043	&	6552.06	&	0.016442	&	21.1231	&	15.2531\\
		&$v_1$=99.043	&	108.12	&	0.916043	&	5075.2	&	0.016442	&	21.1231	&	15.2531\\
		\hline
		\multirow{2}{*}{1.7}
		&$v_2$=127.373	&	135.672	&	0.93883	&	6402.96	&	0.0176299	&	21.2375	&	19.9147\\
		&$v_1$=106.568	&	113.511	&	0.93883	&	5357.1	&	0.0176299	&	21.2375	&	19.9147\\
		\hline
		\multirow{2}{*}{1.8}
		&$v_2$=126.821	&	132.089	&	0.960119	&	6267.14	&	0.0187957	&	21.3509	&	25.4201\\
		&$v_1$=113.432	&	118.144	&	0.960119	&	5605.5	&	0.0187957	&	21.3509	&	25.4201\\
		\hline
		\multirow{2}{*}{1.9}
		&$v_2$=126.226	&	128.789	&	0.980101	&	6142.58	&	0.01994	&	21.4628	&	31.8231\\
		&$v_1$=119.748	&	122.18	&	0.980101	&	5827.37	&	0.01994	&	21.4628	&	31.8231\\
		\hline
		\multirow{2}{*}{2}
		&$v_2$=125.602	&	125.735	&	0.998942	&	6027.68	&	0.0210635	&	21.5728	&	39.1746\\
		&$v_1$=125.602	&	125.735	&	0.998942	&	6027.68	&	0.0210635	&	21.5728	&	39.1746\\
		\hline
		\hline
	   \end{tabular}}
	\caption{Results in case $m_{H^{\pm}}(v_2)=m_A(v_2)=300$ GeV, $m_H(v_2)=150$ GeV.}
\end{table}

\begin{table}[H]
  \centering
\resizebox{0.9\textwidth}{!}{\begin{tabular}{||c||c|c|c|c|c|c|c||}\hline \hline
		$a$&$v_c$ [GeV]&$T_C$ [GeV]&$S$&$E$ [GeV]&$\alpha$&$\betah$&$\Omega h^2 (f_{peak})\times 10^{-13}$ \\
		\hline
		\hline
		\multirow{2}{*}{1.1}
		&$v_2$=198.272	&	171.699	&	1.15477	&	8634.25	&	0.0319842	&	22.6293	&	1.87205\\
		&$v_1$=62.6992	&	54.2959	&	1.15477	&	2730.39	&	0.0319842	&	22.6293	&	1.87205\\
		\hline
		\multirow{2}{*}{1.2}
		&$v_2$=195.78	&	164.759	&	1.18829	&	8377.36	&	0.0346998	&	22.8808	&	2.52906\\
		&$v_1$=87.5557	&	73.6823	&	1.18829	&	3746.47	&	0.0346998	&	22.8808	&	2.52906\\
		\hline
		\multirow{2}{*}{1.3}
		&$v_2$=193.537	&	158.617	&	1.22015	&	8150.04	&	0.037412	&	23.1218	&	3.33497\\
		&$v_1$=106.004	&	86.8783	&	1.22015	&	4463.96	&	0.037412	&	23.1218	&	3.33497\\
		\hline
		\multirow{2}{*}{1.4}
		&$v_2$=191.544	&	153.149	&	1.25071	&	7947.12	&	0.0401542	&	23.3511	&	4.32144\\
		&$v_1$=121.143	&	96.86	&	1.25071	&	5026.2	&	0.0401542	&	23.3511	&	4.32144\\
		\hline
		\multirow{2}{*}{1.5}
		&$v_2$=189.799	&	148.262	&	1.28016	&	7764.6	&	0.0429607	&	23.5669	&	5.53165\\
		&$v_1$=134.208	&	104.837	&	1.28016	&	5490.4	&	0.0429607	&	23.5669	&	5.53165\\
		\hline
		\multirow{2}{*}{1.6}
		&$v_2$=188.289	&	143.888	&	1.30859	&	7599.4	&	0.0458661	&	23.7667	&	7.02374\\
		&$v_1$=145.848	&	111.455	&	1.30859	&	5886.47	&	0.0458661	&	23.7667	&	7.02374\\
		\hline
		\multirow{2}{*}{1.7}
		&$v_2$=186.999	&	139.975	&	1.33595	&	7449.03	&	0.0489043	&	23.9476	&	8.87484\\
		&$v_1$=156.455	&	117.111	&	1.33595	&	6232.31	&	0.0489043	&	23.9476	&	8.87484\\
		\hline
		\multirow{2}{*}{1.8}
		&$v_2$=185.909	&	136.483	&	1.36213	&	7311.46	&	0.0521068	&	24.1066	&	11.1852\\
		&$v_1$=166.282	&	122.074	&	1.36213	&	6539.57	&	0.0521068	&	24.1066	&	11.1852\\
		\hline
		\multirow{2}{*}{1.9}
		&$v_2$=184.994	&	133.381	&	1.38696	&	7184.98	&	0.0555005	&	24.2407	&	14.0822\\
		&$v_1$=175.5	&	126.536	&	1.38696	&	6816.27	&	0.0555005	&	24.2407	&	14.0822\\
		\hline
		\multirow{2}{*}{2}
		&$v_2$=184.227	&	130.638	&	1.41021	&	7068.15	&	0.0591052	&	24.3471	&	17.7223\\
		&$v_1$=184.227	&	130.638	&	1.41021	&	7068.15	&	0.0591052	&	24.3471	&	17.7223\\
		\hline
		\hline
\end{tabular}}
	\caption{Results in case $m_{H^{\pm}}(v_2)=m_A(v_2)=500$ GeV, $m_H(v_2)=250$ GeV.}
\end{table}

\begin{table}[H]
   \centering
\resizebox{0.9\textwidth}{!}{\begin{tabular}{||c||c|c|c|c|c|c|c||}\hline \hline
		$a$&$v_c$ [GeV]&$T_C$ [GeV]&$S$&$E$ [GeV]&$\alpha$&$\betah$&$\Omega h^2 (f_{peak})\times 10^{-12}$ \\
		\hline
		\hline
		\multirow{2}{*}{1.1}
		&$v_2$=226.219	&	173.466	&	1.30411	&	9149.73	&	0.0453932	&	23.7359	&	0.676314\\
		&$v_1$=71.5367	&	54.8549	&	1.30411	&	2893.4	&	0.0453932	&	23.7359	&	0.676314\\
		\hline
		\multirow{2}{*}{1.2}
		&$v_2$=224.01	&	166.748	&	1.34341	&	8891.16	&	0.0497831	&	23.9945	&	0.94705\\
		&$v_1$=100.18	&	74.5718	&	1.34341	&	3976.25	&	0.0497831	&	23.9945	&	0.94705\\
		\hline
		\multirow{2}{*}{1.3}
		&$v_2$=222.28	&	161.054	&	1.38016	&	8663.24	&	0.0545297	&	24.2059	&	1.32034\\
		&$v_1$=121.748	&	88.213	&	1.38016	&	4745.05	&	0.0545297	&	24.2059	&	1.32034\\
		\hline
		\multirow{2}{*}{1.4}
		&$v_2$=220.939	&	156.277	&	1.41376	&	8460.32	&	0.0597024	&	24.3614	&	1.83858\\
		&$v_1$=139.734	&	98.8385	&	1.41376	&	5350.77	&	0.0597024	&	24.3614	&	1.83858\\
		\hline
		\multirow{2}{*}{1.5}
		&$v_2$=219.885	&	152.323	&	1.44354	&	8277.83	&	0.0653321	&	24.4547	&	2.55674\\
		&$v_1$=155.482	&	107.709	&	1.44354	&	5853.31	&	0.0653321	&	24.4547	&	2.55674\\
		\hline
		\multirow{2}{*}{1.6}
		&$v_2$=219.015	&	149.102	&	1.4689	&	8112.05	&	0.0713998	&	24.4827	&	3.54026\\
		&$v_1$=169.648	&	115.494	&	1.4689	&	6283.56	&	0.0713998	&	24.4827	&	3.54026\\
		\hline
		\multirow{2}{*}{1.7}
		&$v_2$=218.235	&	146.522	&	1.48943	&	7959.9	&	0.0778348	&	24.4464	&	4.85909\\
		&$v_1$=182.588	&	122.589	&	1.48943	&	6659.73	&	0.0778348	&	24.4464	&	4.85909\\
		\hline
		\multirow{2}{*}{1.8}
		&$v_2$=217.466	&	144.494	&	1.50501	&	7818.9	&	0.0845259	&	24.3505	&	6.57865\\
		&$v_1$=194.508	&	129.24	&	1.50501	&	6993.43	&	0.0845259	&	24.3505	&	6.57865\\
		\hline
		\multirow{2}{*}{1.9}
		&$v_2$=216.653	&	142.93	&	1.5158	&	7687.09	&	0.0913395	&	24.2019	&	8.74961\\
		&$v_1$=205.535	&	135.596	&	1.5158	&	7292.61	&	0.0913395	&	24.2019	&	8.74961\\
		\hline
		\multirow{2}{*}{2}
		&$v_2$=215.759	&	141.75	&	1.52211	&	7562.95	&	0.0981397	&	24.0094	&	11.3997\\
		&$v_1$=215.759	&	141.75	&	1.52211	&	7562.95	&	0.0981397	&	24.0094	&	11.3997\\
		\hline
		\hline
\end{tabular}}
	\caption{Results in case $m_{H^{\pm}}(v_2)=m_A(v_2)=600$ GeV, $m_H(v_2)=300$ GeV.}
\end{table}

\end{document}